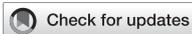

# Observation of critical scaling in spin glasses below $T_c$ using thermoremanent magnetization

G. G. Kenning[1]*, M. Brandt[1], R. Brake[1], M. Hepler[1] and D. Tennant[2]

[1]Madia Department of Chemistry, Biochemistry, Physics and Engineering, Indiana University of Pennsylvania, Indiana, PA, United States, [2]Texas Materials Institute, The University of Texas at Austin, Austin, TX, United States

Time-dependent thermoremanent magnetization (TRM) studies have been instrumental in probing energy dynamics within the spin glass phase. In this paper, we review the evolution of the TRM experiment over the last half century and discuss some aspects related to how it has been used in the understanding of spin glasses. We also report on recent experiments using high-resolution DC SQUID magnetometry to probe the TRM at temperatures less than but near to the transition temperature $T_c$. These experiments have been performed as a function of waiting time, temperature, and five different magnetic fields. We find that as the transition temperature is approached from below, the characteristic time scale of TRM is suppressed up to several orders of magnitude in time. In the highest-temperature region, we find that the waiting time effect subsides, and a waiting time-independent crossover line is reached. We also find that increasing the magnetic field further suppresses the crossover line. Using a first-principles energy argument across the crossover line, we derive an equation that is an excellent fit to the crossover lines for all magnetic fields probed. The data show strong evidence for critical slowing down and an *H* = *0 Oe* phase transition.

## 1 Introduction

The main goal of this paper is to present the data and analysis which use the thermoremanent magnetization (TRM) waiting time effect in spin glasses as a probe of the critical region near the spin glass phase transition. In the spirit of this collection, we begin with a brief historical perspective and a primer on several magnetic signatures found in spin glasses. This introductory section includes a description of the experiments, field cooled/zero field cooled (FC/ZFC) and field cooled-thermoremanent magnetization/zero field cooled magnetization (FC-TRM/ZFC-TRM), the waiting time effect, and the relationships between them. We then review the structure of the FC-TRM decay and discuss several experiments and simulations that are important for understanding the data and the analysis to follow. This is not meant to be a comprehensive review. Since the measurements presented in this paper represent an improvement in experimental design and an improved signal-to-noise ratio, Section 2, *Experimental methods*, is more detailed and may be of use to experts in the field. Finally, in Section 3 and Section 4, we present the data and analysis which encompass using the TRM and waiting time effect in spin glasses as a probe of the critical region near the spin glass phase transition.

The history of experimental and numerical studies in spin glasses follows the development of technology itself, and it can be argued that the challenges of exploring the nature of the spin glass phase have driven aspects of technology forward. Early numerical studies on this NP hard problem [1] began in the 1980s with simulations on smaller than 16 spins. Today, with the advent of large-scale computing, dedicated computers, and advances in algorithms, simulations are done with more than $1.67x10^7$ spins [2]. Even as we reach toward large-scale neural networks [3] and quantum computing [4], the spin glass algorithm is seen as a fundamental starting point.

From the experimental side, in small magnetic fields, the spin glass state has, by its random nature, a small magnetic signal. Elucidating the much smaller time dependencies of the magnetization signals makes the measurement sensitivity of primary importance. These signals approach zero at several limits, ($H \to 0\ Oe$, $T \to T_g$, and $t \to \infty$), making the signal-to-noise ratio crucial in investigations of these limits. Over the last five decades, spin glass experiments have increased orders of magnitude in their sensitivity, stability, and time scales. From direct Faraday-effect measurements to AC measurements with lock in amplifiers, RF SQUIDS, and DC SQUIDS, sensitivity has increased, and new regions of the spin glass state have become accessible. The PC control of the experiments now allows us to perform experiments that were not possible before 1980. We can now perform many automated experiments with temperatures controlled to the ± μK resolution for long periods of time.

The first TRM measurements were made by [5], using Faraday techniques, only 2 years after the discovery of the spin glass phase [6]. This measurement was made by cooling the spin glass below its transition temperature, in a magnetic field, to a measuring temperature. These measurements were often made by physically pulling the sample out of a sensing coil, thereby inducing a magnetic flux change in the coil and then electronically integrating the signal to obtain total magnetization. Other "static" measurements then evolved, including the first field-cooled/zero field-cooled (FC/ZFC) measurements. They were performed on $Gd_{0.37}Al_{0.63}$ [7] and shortly afterward on the "canonical" spin glasses AuFe [8] and CuMn [9], cementing the FC/ZFC magnetization curves as a spin glass signature. This measurement became a quick way of determining the approximate spin glass transition temperature with both a peak in the ZFC and the onset of irreversibility occurring at a temperature approximately equal to the peak temperature of the low-frequency AC susceptibility.

To understand the FC-TRM decay measurement, the corresponding ZFC-TRM measurement, and the subtle differences between them, it is useful to analyze the FC/ZFC magnetization of the spin glass state as a function of temperature. Figure 1 displays the field cooled (FC) and zero field-cooled (ZFC) magnetization curves for a poly-crystalline $Cu_{0.95}Mn_{0.05}$ sample. These measurements were taken using the Quantum Design DC SQUID MPMS magnetometer at the University of Texas. To produce this curve, one starts in a zero magnetic field, at a high temperature (in this case, 40 K), above the spin glass transition temperature (in this case, approximately 28.7 K). The sample is then rapidly cooled down to a temperature below the spin glass transition temperature (in this case, 10 K). A field of 10 Oe is then applied. The magnetization makes a rapid jump of approximately $6x10^{-5}$ emu, and the magnetization value is read. The temperature is incremented (in this case, by 0.5 K), and at each point up to the maximum temperature, the magnetization is recorded. This curve is the ZFC curve. Starting from that same high temperature in the same field (10 Oe), the temperature is then lowered in similar temperature increments and magnetization measured at each temperature all the way down to the lowest temperature. This is the FC magnetization curve.

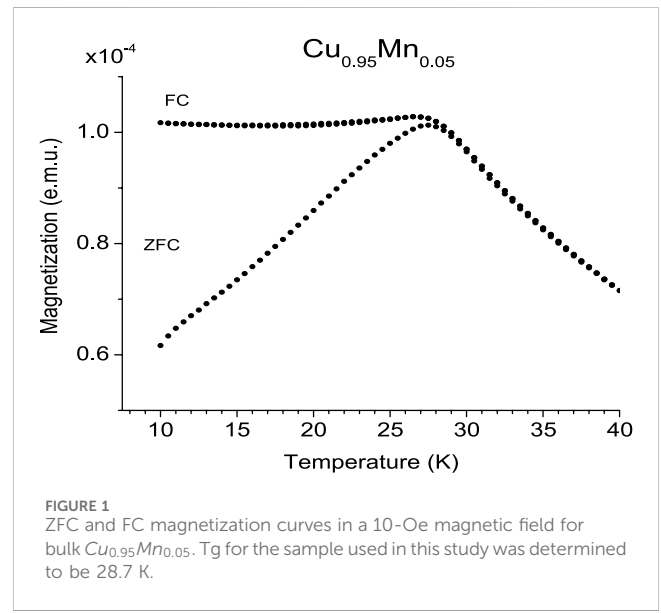

FIGURE 1
ZFC and FC magnetization curves in a 10-Oe magnetic field for bulk $Cu_{0.95}Mn_{0.05}$. Tg for the sample used in this study was determined to be 28.7 K.

One problem that arises in the experimental spin glass field is that unlike numerical studies, which determine $T_c$ from the Binder cumulant, it is difficult to experimentally determine the actual zero-field spin glass phase transition temperature $T_c$. Perhaps, the best attempt was made by [10], who extrapolated the diverging terms in non-linear susceptibility. This, however, is a difficult measurement to perform, and more rapid approximations of $T_g$ have been made from the FC/ZFC or AC measurements. In this paper, we call the experimental approximation of the phase transition temperature, $T_g$. In this collection, [11], compare the values of $T_g$ obtained from different techniques and showed the limitations to these methods. In this paper, we use the $H \to 0\ Oe$ extrapolated value of the onset of irreversibility to determine a value of $T_g = 28.7\ K$ for the $Cu_{0.95}Mn_{0.05}$ polycrystalline sample measured in Figure 1 and $T_g = 31.5\ K$ for the single-crystal $Cu_{0.94}Mn_{0.06}$ sample used in all other experiments in this paper.

Getting back to the FC/ZFC curves, above $T_g$, the FC and ZFC magnetization overlap displaying the reversible Curie–Weiss behavior of a paramagnetic phase. At a temperature near the transition temperature, FC magnetization becomes approximately constant and remains close to that value throughout most of the spin glass phase. At first observation, it looks like the spin glass magnetization (and spin configuration) freezes at Tg. It can be observed that there is a peak in the ZFC curve, which is often used as a determination of $T_g$ [12]. It can also be observed that near the peak temperature, the FC and ZFC curves split, indicating the onset of magnetic irreversibility in the system. This onset temperature has also been used to determine the transition temperature [13]. The peak in the ZFC magnetization appears to be field-independent, while the onset of irreversibility has a definite magnetic field dependence, weakly shifting down in temperature as the magnetic field increases. The ZFC-TRM and the FC-TRM both measure time dependencies of this irreversibility.

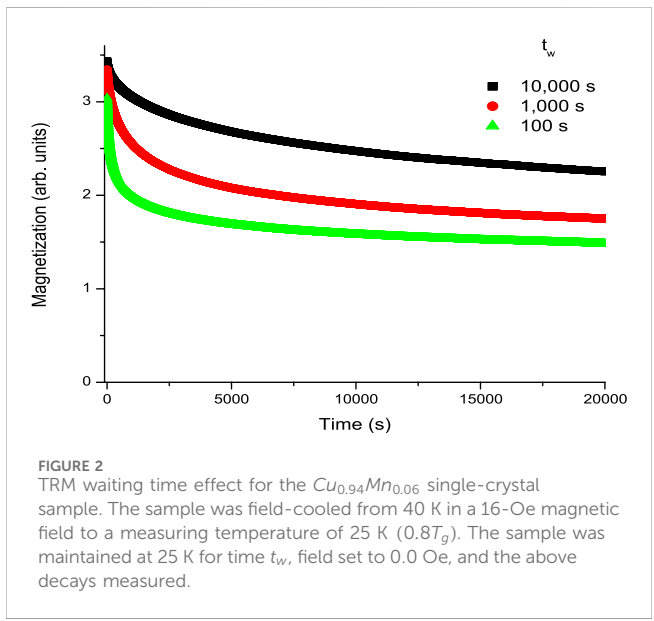

FIGURE 2
TRM waiting time effect for the $Cu_{0.94}Mn_{0.06}$ single-crystal sample. The sample was field-cooled from 40 K in a 16-Oe magnetic field to a measuring temperature of 25 K ($0.8T_g$). The sample was maintained at 25 K for time $t_w$, field set to 0.0 Oe, and the above decays measured.

In 1983, [14] reported both a time-dependent decay of the ZFC-TRM and the waiting time effect using RF-SQUID magnetometry. In the ZFC-TRM, the sample is cooled in a zero magnetic field from a high temperature above the transition temperature to some measuring temperature within the spin glass state. Since the sample is cooled in a zero magnetic field, the random nature of the spin glass state implies zero magnetization (time reversal symmetry applies). In this zero-magnetization state, the sample is maintained at the measuring temperature for time $t_w$, after which the magnetic field is turned on and the magnetization measured at small time intervals for a measuring time $t_m$. On application of the magnetic field, there is a rapid magnetization jump to a value approximately equal to the ZFC magnetization at that temperature (with a small systematic difference due to the waiting time effect), and then, a slow increase occurs. This increase is the ZFC-TRM. Previously, it was generally assumed that the increase would come to equilibrium magnetization at the FC line; however, this has been called into question by the weak logarithmic time dependencies of the FC magnetization [15, 16].

In 1984, [17] reported similar time dependencies (including the waiting time effect) in RF-SQUID-aided measurements of the FC-TRM measurement. In the FC-TRM measurement, the sample is cooled in a magnetic field, from a temperature above the transition temperature to a measuring temperature in the spin glass state. This is the same procedure as the FC magnetization measurement (Figure 1). Therefore, at the measuring temperature, the system starts out with a magnetization equal to the FC magnetization (with the small deviation due to the weak logarithmic time-dependent decay of the FC magnetization) [15, 16]. In this magnetized state, at a constant temperature $T_m$, the sample is held for a time $t_w$. After this, the magnetic field is turned off, and magnetization is measured at small time intervals for a measuring time $t_m$. After the magnetic field is shut off, there is a rapid decrease in magnetization by an amount approximately equal to the ZFC magnetization at that temperature (with a small systematic difference due to the waiting time effect), and then, a slow decay occurs. This decay is the FC-TRM. From symmetry arguments, the final equilibrium magnetization, in the zero field, must be zero. Figure 2 displays the data observed in the FC-TRM measurement. The waiting time effect is clear in the shift of the curves, but there are other subtle differences between curves, which became clearer with further analysis.

Over most of the temperatures below the spin glass transition temperature, superposition appears to hold and can be described by Equation 1 [18]

$$M_{TRM}(t_w, t) = M_{FC}(0, t + t_w) - M_{ZFC}(t_w, t). \quad (1)$$

As mentioned previously, $M_{FC}(0, t + t_w)$ has a weak time dependence, but comparatively, it is small enough that its effect on the above equation can be ignored.

Superposition assumes that the removal of the magnetic field in the FC-TRM measurement is equivalent to adding a negative field to the sample in the FC state at time $t_w$. The implications of this are that (for reasonably small fields) the absolute value of the magnetic field does not matter, and only the change in magnetic field is important. This implies that the manifold of states into which the spin glass state freezes is effectively equivalent for small (<100 G) fields, including the zero magnetic field. Recently, it was found that the superposition principle breaks down close to $T_g$, and as the magnetic field increases, it may only be valid as $H \to 0$ [19].

An early attempt to analyze the entire decay curve was made using a stretched exponential function, Equation 2 [17].

$$M_{TRM}(t_w, t) = M_0 \lambda^{-\alpha} exp\left[-\left(\frac{t}{\tau_p}\right)^{1-n}\right]. \quad (2)$$

The power law was later added to describe the short time (<1 s) rapid decrease in magnetization [20]. $\lambda$ is a an effective time scale that, in the short time limit $t \ll t_w$, $\lambda \simeq t$, which is waiting time-independent. For that reason, this term is often called the stationary term [21]. Finally, regarding the structure of the entire TRM decay, there is a third time regime. It has been observed, using short waiting time and long measurement time scales, that the waiting time effect has a finite lifetime [22]. Using a fast cooling protocol [23], TRM decay measurements with short waiting times (7–100 s), measured over long times 10,000–100,000 s, show that the waiting time-dependent part of the total decay ends, and curves with different waiting times converge into a single waiting time-independent logarithmic decay. At low temperatures, this logarithmic term (for accessible waiting time) dominates the irreversibility, but the relative magnitude (log decay/waiting time decay) decreases as the transition temperature is approached [24]. The longer the waiting time, the further the decay, extending into the logarithmic term, strongly implying that the waiting time effect and the logarithmic term are related.

It was found that over most of the spin glass state that the stretching exponent n was a constant (see [25]; Figure 2). As $T_g$ was approached, this constant deviated toward n = 1. In order to fit this function, the time scale $\tau_0$ changed by approximately eight orders of magnitude (see [25]; Figure 3). The lack of a first-principles theory that predicts a stretched exponential has led to a decrease in the use

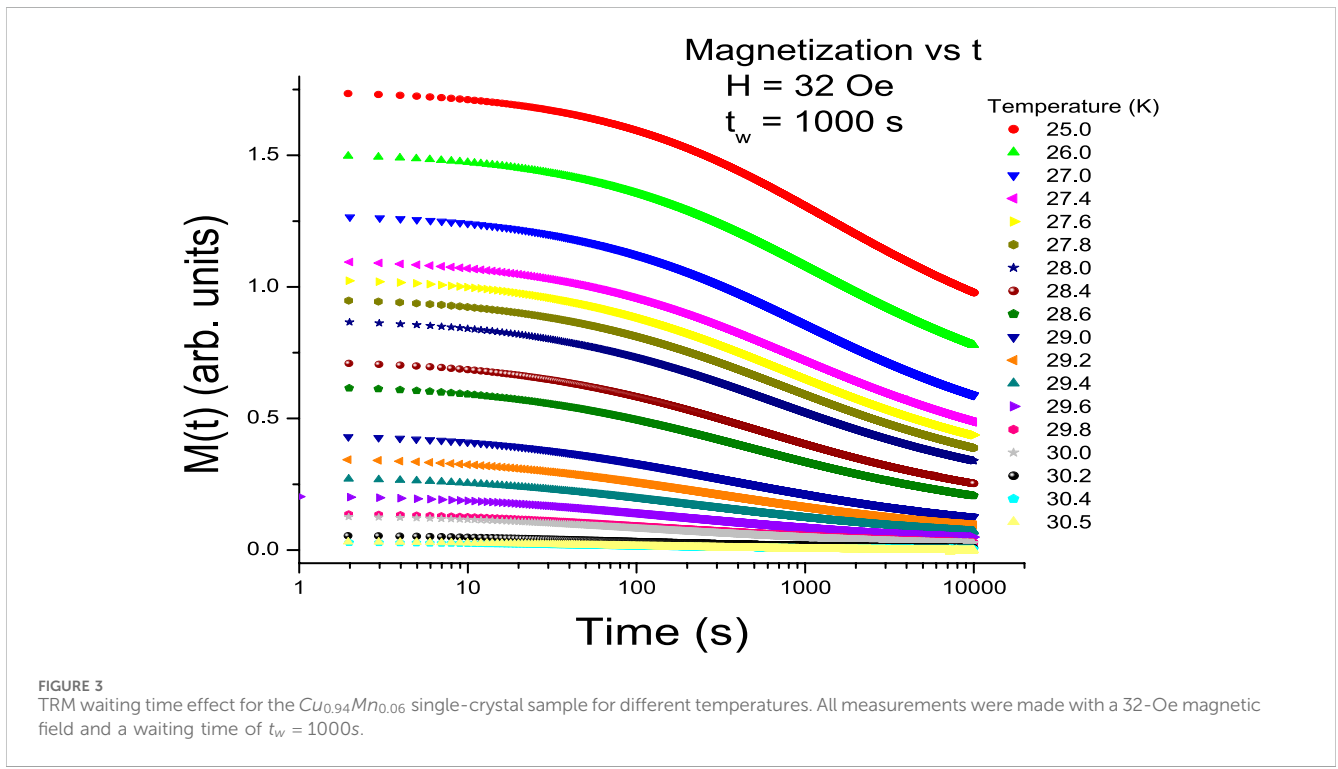

FIGURE 3
TRM waiting time effect for the $Cu_{0.94}Mn_{0.06}$ single-crystal sample for different temperatures. All measurements were made with a 32-Oe magnetic field and a waiting time of $t_w = 1000s$.

of the function, although it is still used by some spin glass researchers.

A second analysis then evolved, where [26, 27] applied the phenomenological time scaling technique (first used by [28] to quantify the waiting time effect in polymers) to spin glasses. In this technique, all curves produced with different waiting times could be collapsed onto a single master curve by scaling the data with a reduced effective time scale defined by Equations 3, 4.

$$\xi = \lambda / t_w^\mu, \quad (3)$$

where

$$\lambda = \frac{t_w}{1-\mu}\left[\left(1 + \frac{t}{t_w}\right)^{1-\mu} - 1\right], \quad \mu < 1. \quad (4)$$

In the limit $t \ll t_w$, $\xi$ reduces to $\frac{t}{t_w^\mu}$.

A new parameter $\mu$ is introduced. This proposes that all TRM decays, at a fixed temperature, with a wide range of waiting times can be effectively collapsed with a single parameter. This type of scaling is referred to as $\mu$-scaling. In [27] (Figure 2—inset), it can be observed that over most of the spin glass phase, $\mu$ is approximately constant. Sub-aging is observed over most of the spin glass phase with $\mu$ approximately equal to 0.9. From $0.8T_g \rightarrow 1.0T_g$, $\mu$ decreases. In a follow-up study, [29] found that at approximately $0.96T_g$, the decay curves could be collapsed at long measuring times or short measuring times but not both. This result suggests that $\mu$ scaling does not hold as the transition temperature is approached. Much of the impetus for the experimental work to follow comes from observations of the stretched exponential and $\mu$ scaling analysis and other deviations observed as the transition temperature is approached.

Perhaps, the most interesting aspect of the waiting time effect is that it appears to be effectively temperature-independent in the approximate temperature range $0.4T_g$–$0.8T_g$. Not only are the n values (from the stretched exponential) and $\mu$ values (from $\mu$ scaling) approximately constant in this region but the magnitude of the waiting time-dependent magnetization decay is also approximately constant over this temperature regime. This is rather amazing as the decays are fundamentally governed by thermally activated dynamics, which, in most cases, leads to Arrhenius behavior and very large temperature dependencies. A third method of analyzing the TRM data was put forth by [30]. They observed that on a logarithmic scale, the TRM decay displays an inflection point (see, for example, Figure 3). By plotting the logarithmic derivative Equation 5.

$$S(t) = -\frac{dM(t)}{dln(t)}, \quad (5)$$

the S(t) function displays a peak at a time equal to the time where the inflection point in the decay is observed. This time is called $t_w^{eff}$. Figure 4A displays the S(t) function for the 32-Oe magnetization data shown in Figure 3. The highest temperature data are shown in Figure 4B to show experimental resolution.

The S(t) function, as well as the associated characteristic time scale, $t_w^{eff}$ is a straightforward method of assigning a single parameter associated with the waiting time effect. In the temperature range $0.4T_g$–$0.8T_g$, this characteristic time scale is observed to occur at a time approximately equal to the input waiting time. The implication of this is that the decay reflects the time scale input during the waiting time. Actually, in the above mentioned temperature range, the S(t) function has a peak closer to $2t_w$, leading some researchers to use the total time $(t + t_w)$ as the correct time, where t is the measuring time. In this sense, the correct time is the total time spent at the measuring temperature. In this paper, we use the S(t) function to investigate time and spatial dependencies in the spin glass state near $T_g$, in particular focusing on the region $T > 0.9 T_g$.

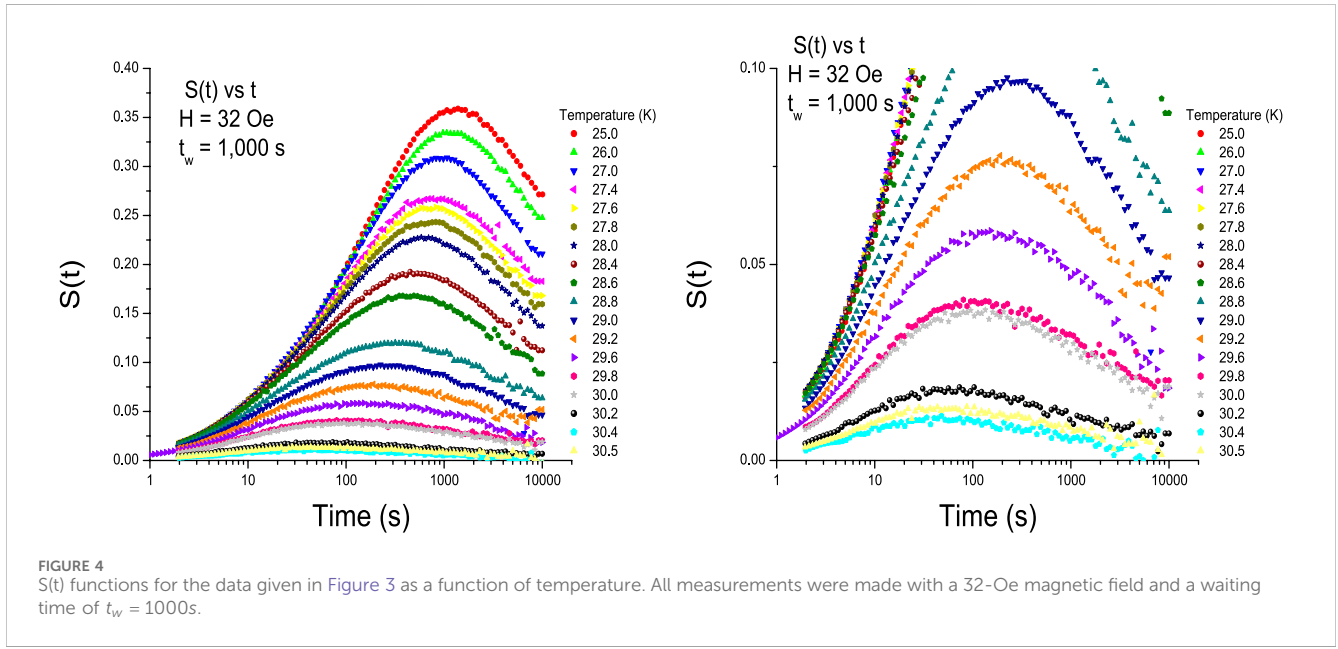

FIGURE 4
S(t) functions for the data given in Figure 3 as a function of temperature. All measurements were made with a 32-Oe magnetic field and a waiting time of $t_w = 1000 s$.

[31] analyzed 2D and 3D numerical simulations of Ising spin glass models. They found that they could determine a spatially dependent coherence length scale using a 4-spin autocorrelation function, Equation 6.

$$G_T(r, t_w) = \frac{1}{N} \sum_{i=1}^{N} \frac{1}{t_w} \sum_{t=t_w}^{2t_w-1} \left[ \langle S_i^a(t) S_{i+r}^a(t) S_i^b(t) S_{i+r}^b(t) \rangle \right]_{av}. \quad (6)$$

This spatial coherence length is observed to grow as a power law according to Equation 7.

$$\xi(t_w, T) = c_1 \left( \frac{t_w}{\tau_0} \right)^{c_2(T/T_c)}, \quad (7)$$

where $\tau_o$ is a microscopic exchange time and $c_1$ and $c_2$ are constants.

This dynamic analysis was extended to CuMn(14%) thin films by [32], who found consistent results for three films with substantially different $\mathcal{L}$, using $c_1 = 1.448$ and $c_2 = .104$. They also associated the maximum barrier with the observed thin-film freezing temperatures $\Delta_{max} = k_b T_f(\mathcal{L})$ and found that Equation 7 substantially predicts the form of $T_f(\mathcal{L})$. In thin films, the FC/ZFC signatures look similar to bulk samples but at lower temperatures. We call the approximate freezing temperature in thin films $T_f$ to discriminate it from the bulk value $T_g$. In the above studies, it is assumed that $\xi$ grows isotropically until it reaches the thickness of the film. Slightly different results were found ($c_1 = 0.87$ and $c_2 = 0.11$) [33] using the freezing temperature associated with finite size effects in spin glass films [34].

Numerical analysis of the 4-spin correlation function showed that the power law growth (Equation 7) holds right up to the spin glass transition temperature, at which point the exponent $c_2(T/T_c)$ becomes a constant [2]. This is quite different from what we observed for the single-crystal sample in 16 Oe (Figure 5). Using $t_w^{eff}$ in Equation 7 for $t_w$, it is observed that the calculated coherence length first grows as a power law [33] and then decreases as $T_g$ is approached.

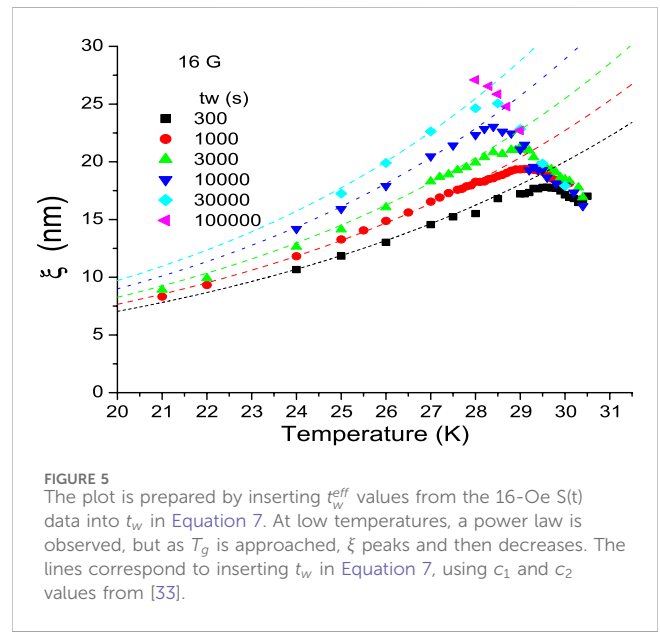

FIGURE 5
The plot is prepared by inserting $t_w^{eff}$ values from the 16-Oe S(t) data into $t_w$ in Equation 7. At low temperatures, a power law is observed, but as $T_g$ is approached, $\xi$ peaks and then decreases. The lines correspond to inserting $t_w$ in Equation 7, using $c_1$ and $c_2$ values from [33].

Figure 6 shows the entire magnitude of the observed remanent magnetization for the single-crystal sample measured using a TRM protocol with a 16-Oe field. The solid symbols on this graph represent the remanent magnetization signal a few seconds after the magnetic field is shut off. This signal is comparable to the remanence defined by the difference in FC and ZFC magnetization. The open symbols are the corresponding measurement of the decay 10,000 s after the magnetic field was switched to $H = 0$ Oe. For the shorter waiting time (i.e., 1,000 s), the waiting time effect is almost over at 10,000 s, and the remaining magnetization (i.e., below the open circles) decays logarithmically, as discussed previously. This decay happens on time scales much larger than the experimental time scales reported in this study.

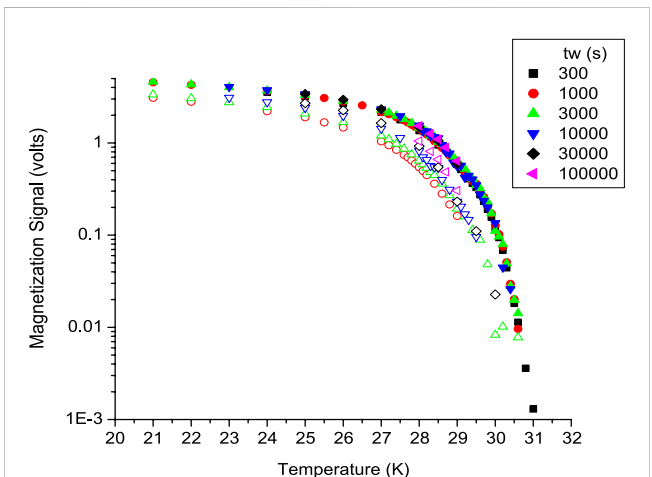

FIGURE 6
Magnitude of the observed remanent magnetization for the single-crystal sample cooled in a 16-Oe magnetic field. The solid symbols on this graph represent the remanent magnetization signal 12 s after the magnetic field is shut off. This signal is comparable to the remanence defined by the difference in the FC and ZFC magnetizations. The open symbols are the corresponding measurement of the decay, 10,000 s after the magnetic field was switched to H = 0 Oe. For the shorter waiting time (i.e., 1,000 s), the waiting time effect is almost over at 10,000 s, and the remaining magnetization (i.e., below the open circles) decays logarithmically, as discussed previously.

Scaling theory and the underlying renormalization group theory have opened a path for understanding critical phenomena near phase transitions [35]. As the critical temperature of a phase transition is approached (either from high or low temperatures), the physics of the system is governed by the correlated growth of fluctuations and critical decreasing of fluctuation time scales. Near the thermodynamic critical point of a continuous magnetic phase transition, strong highly correlated magnetization fluctuations are expected that, in principle, can occur with any time and/or length scale [36]. For the discussion to follow, we plot the critical fluctuation time scale as a function of temperature, Figure 7A, and the critical correlation length scale, Figure 7B, as a function of temperature. In Figure 7A, we plot $\tau_c$ vs. T, Equation 8. The value for $\tau_{oc}$ corresponds to a transition temperature of 32.4 K ($\tau_{oc} = \frac{h}{k_B T_c} = 1.48 x 10^{-12}$ s).

$$\tau_c = \tau_{oc} \left| \frac{T - T_c}{T_c} \right|^{-z\nu} . \qquad (8)$$

In Figure 7B, we plot $\xi_c$ vs. T, Equation 9, with $\xi_{oc} = 2.8 x 10^{-10}$ nm, where $\xi_{oc}$ is the mean nn Mn distance in $Cu_{0.94}Mn_{0.06}$.

$$\xi_c = \xi_{oc} \left| \frac{T - T_c}{T_c} \right|^{-\nu} . \qquad (9)$$

These plots will be useful for understanding the theory proposed in Section 4.

## 2 Experimental methods

The challenge for TRM measurements close to, but below the transition temperature, is the need for extreme sensitivity. To begin with, the total remanent magnetization (and hence, the signal) rapidly decreases as the transition temperature is approached. This can be observed in Figure 6. In addition, the S(t) function is a derivative, significantly enhancing the noise observed in the magnetization decay. To probe close to the transition temperature, we build a dedicated, very high-sensitivity dual-DC SQUID magnetometer. The Indiana

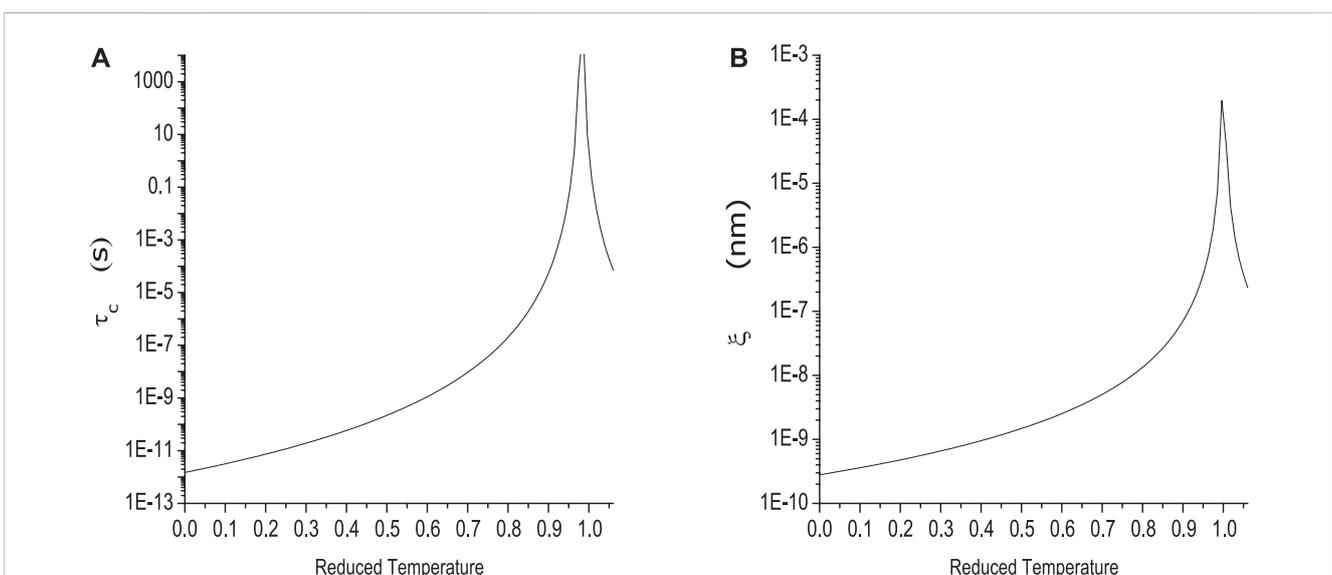

FIGURE 7
In (A) (top), we plot $\tau_c$ from Equation 8 with $\tau_{oc} = 1.48 x 10^{-12}$ s, using $T_c = 32.4$ K. In (B) (bottom), we plot $\xi_c$ from Equation 9 with $\xi_{oc} = 2.8 x 10^{-10}$ nm. These graphs are plotted on a reduced scale ($T/T_c$).

University of Pennsylvania (IUP) magnetometer has two independent SQUID amplifiers. One measures the sample, and the other measures the ambient background. The pickup coils have nearly identical second-order gradiometer configurations and are displaced from each other by 10 cm. The pickup coils have a diameter of 1.1 cm. The resolution of the magnetometer at the baseline point is ±10 nano-emu. This is significantly better than the University of Texas Quantum Design DC SQUID magnetometer, which has a resolution of >±50 nano-emu. A comparison with the Quantum Design MPMS DC SQUID magnetometer is given in [37]. In addition we lowered the noise floor by a factor of five through enhanced pressure control of the helium bath. More details on the IUP magnetometer are given in [37].

The TRM measurements reported here are an analog measurement. The details are briefly discussed. In a magnetic field, the sample is brought from a high temperature, usually 5–6 K above $T_g$, down to the measuring temperature below $T_g$. The temperature reaches $T_m \pm 1 mK$ within the first 100 s, so with significantly longer waiting times, we can ensure that the experiment is highly isothermal. The sample is held at the measuring temperature for a time $t_w$ before the magnetic field is cut. After the field is cut, we wait approximately 12 s (to ensure that the heaters used to reset the SQUID and pickup coils have cooled) and then start measuring. The first magnetization point measured is the value of $M_o$ plotted in Figure 6 (solid symbols). The DC SQUID continually measures the pickup coil signal, and a digital readout is taken at 1s intervals over the entire measurement. At the end of the measurement, we take a baseline measurement to put the TRM decays on an absolute scale. The baseline is measured in the following way. After the TRM decay is measured (in a zero magnetic field), the sample is raised in temperature to well over the transition temperature (40 K). In the paramagnetic state at zero field, the sample will have zero magnetic moment. The sample is then cooled back to the measuring temperature (in the zero magnetic field), the temperature is stabilized, and after 5 min, a baseline measurement is taken. This yields an absolute magnetization signal (zero magnetization). The data presented have this baseline signal subtracted from the TRM signal, providing an absolute magnetization scale. In this report, for example, in Figure 2, the magnetization scale reported as arb. units is actually volts, taken directly from the DC SQUID amplifier. As always, SQUIDS are susceptible to large jumps in the signal, which we call SQUID jumps. These look like step functions in the analog data and are generally much larger than the TRM signal and noise. They are therefore rather easy to remove post-processing. Having two SQUIDS allows us to observe which SQUID jumps are system-wide and which are confined to a particular SQUID.

To further enhance the stabilization of the system, measurements presented in this paper were made in two experimental sessions over which the magnetometer was kept cold. The data taken with a TRM field of 16 Oe were measured over a period of approximately 4 months in 2021, and the other magnetic field data presented were obtained in a 3-month session in 2022. Long sessions cold, enhanced the thermal stabilization of the equipment. The TRM measurements are isothermal measurements, so temperature control is very important. The sample, located at the end of a temperature-controlled $Al_2O_3$ rod, is centered in one of the pickup coils. Temperature control at the measuring temperature was stable to at least ≤±1 mK during the first month of a session and improved (after a month) to < ± 80$\mu K$ over runs as long as $1x10^5 s$. Squid jumps effectively disappeared after the first month.

The bulk polycrystalline CuMn sample (Figure 1) was prepared by alloying high-purity Cu and Mn and then annealing at high temperature to randomize the Mn within the sample, followed by a rapid thermal quench to 77 K. For many years, it was believed within the spin glass experimental community that to correctly produce these types of metallic spin glasses, (i.e., a bulk CuMn sample), high-purity Cu and Mn must be alloyed and then annealed at high temperature (≈800 K for CuMn for up to 48 h). This was done to randomize Mn within the sample, followed by a rapid thermal quench (in our studies, to 77 K) to lock in the random disorder. This technique, however, produced small crystallites, and the effects of the crystallites on the time dependencies were unknown. In a previous study [37], we found that the time associated with the peak in the S(t) function ($tw_{eff}$) dramatically decreased above $0.9T_g$. It was conjectured that the correlation length may reach the polycrystalline size scale, and the dramatic decrease may be due to finite size effects [38]. In 2016, we began working with single crystals of CuMn in order to eliminate any effects due to crystallite sizes. Deborah Schlagle at Ames Laboratory produced three CuMn single-crystal boules of different Mn concentrations. Cu and Mn were arc-melted several times in an argon environment and cast in a copper mold. The ingot was then processed in a Bridgman furnace. X-ray fluorescence (XRF) and optical observation showed that the beginning of the growth is a single phase. Further details on the production and analysis of the sample, including X-ray diffraction, are presented elsewhere [32, 39]. To date, we found no differences in any of the spin glass signatures between the single-crystal samples (including FC-ZFC, FC-TRM, and ZFC-TRM, waiting time effect, and AC susceptibility) and polycrystalline samples, indicating that crystallite size effects in a polycrystalline spin glass sample have little or no effect and that the technique for producing the single crystals sufficiently randomizes the Mn.

## 3 Data and analysis

Previous studies on $Cu_{0.94}Mn_{0.06}$ single-crystal samples at 16 Oe found that the characteristic time scale ($t_w^{eff}$) associated with the peak of the S(t) function displayed a remarkable decrease as $T_g$ (31.5 K) is approached from temperatures below $T_g$ [33]. These data are shown in Figure 8B. In this study, we extend our measurements to other magnetic fields including 9.6, 16, 32, 64, and 96 Oe. Before we move on to discuss the effects of changing magnetic fields, let us review what we observed with 16 Oe. Like all waiting time effect measurements observed to date on CuMn spin glasses, we observe that below approximately 0.8 $T_g$, the peaks in the S(t) functions ($t_w^{eff}$) approach a value of ≈ $2t_w$. Above 0.8 $T_g$, we find a rapid decrease in both the remanent magnetization and the characteristic time scale $t_w^{eff}$ up to 0.96 $T_g$, where we lose the signal (Figures 3, 6). However, by approximately 29 K, the S(t) peak positions, for different waiting times, are indistinguishable, within the experimental resolution. This result is extremely interesting as it appears to define a cut-off time scale. For waiting times less than this time scale, one observes waiting time-dependent decays, albeit with smaller $t_w^{eff}$ values, i.e., a suppressed waiting time effect. For waiting times larger than the cut-off time scale, the observed characteristic time scale was limited to the cut-off time scale, and therefore, no variation with waiting time is observed.

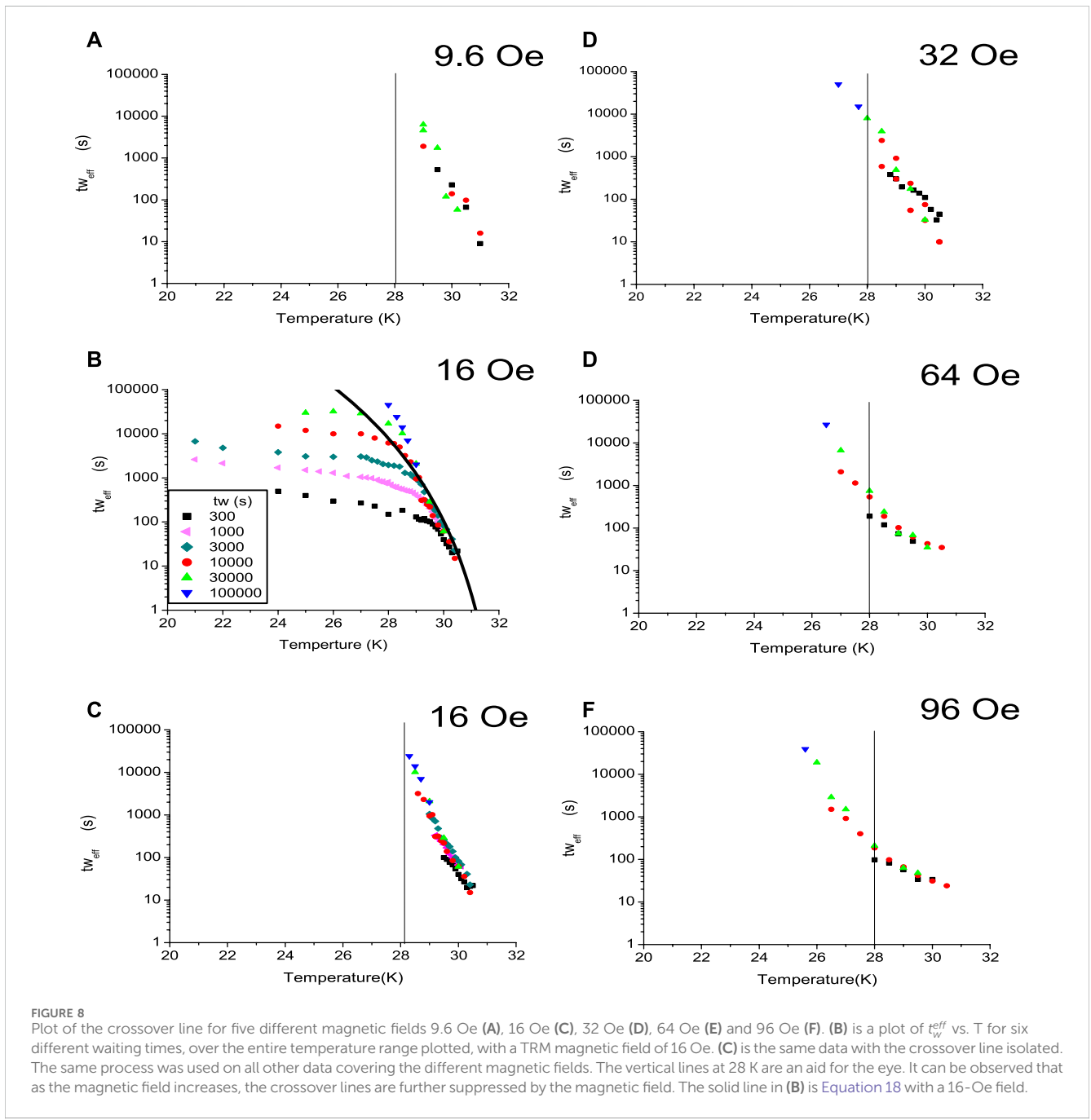

FIGURE 8
Plot of the crossover line for five different magnetic fields 9.6 Oe **(A)**, 16 Oe **(C)**, 32 Oe **(D)**, 64 Oe **(E)** and 96 Oe **(F)**. **(B)** is a plot of $t_w^{eff}$ vs. T for six different waiting times, over the entire temperature range plotted, with a TRM magnetic field of 16 Oe. **(C)** is the same data with the crossover line isolated. The same process was used on all other data covering the different magnetic fields. The vertical lines at 28 K are an aid for the eye. It can be observed that as the magnetic field increases, the crossover lines are further suppressed by the magnetic field. The solid line in **(B)** is Equation 18 with a 16-Oe field.

To begin the analysis of this effect, we first separated the waiting time-independent crossover line from the standard waiting time effect. This was accomplished by removing data associated with the break toward the standard waiting time effect. For example, both Figures 8B,C are plots of the same data (H = 16 Oe), with the waiting time-dependent data removed in Figure 8C. Unfortunately, the exact temperature limits on either side of the crossover line are somewhat subjective.

## 4 Discussion

A model has emerged for the waiting time effect within the spin glass state, and we apply this model to the critical region (near $T_g$). When the sample is cooled in a magnetic field to a measuring temperature within the spin glass state, the magnetization of the entire sample is $M_{fc}$. During the waiting time, there is a growth of coherent regions within the spin glass state, each with a magnetization of $M_{fc}$. Due to a "stiffening" of these coherent regions during the waiting time, these regions are resistant to the change in the magnetic field that occurs after waiting time $t_w$ in the field cooled state and, therefore, decay slowly. As the temperature is increased toward the transition temperature, remanence rapidly decreases, and the effective time scales decrease.

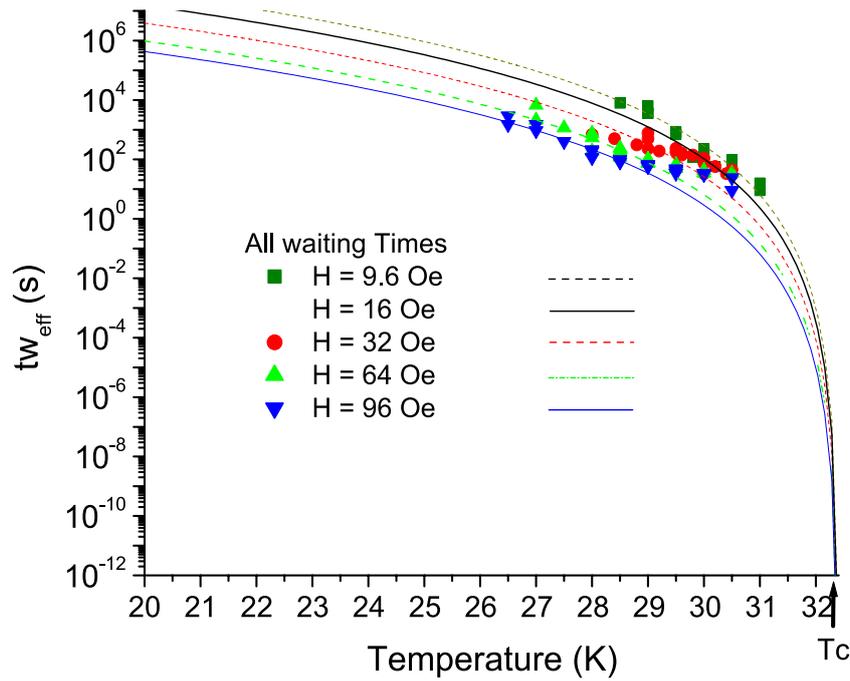

FIGURE 9
The crossover line data are plotted for four different magnetic fields. Plot of Equation 18 (lines) for five different fields. The 16-G data are fit to Equation 18 in Figure 8B.

From this point forward in the analysis, we use $T_c$ instead of $T_g$ to separate the experimental determination of the spin glass transition temperature $T_g$ (approximated from FC/ZFC measurements) from $T_c$, the phase transition fixed point. To understand the crossover line, we treat the energy associated with the waiting time effect as a small perturbation of the total free energy associated with the entire spin glass state below $T_c$. This is plausible as the total remanent magnetization in the temperature region of the crossover line (28 K–31 K) is only a few % of the total FC magnetization (Figure 1). Within the spin glass phase, during the waiting time in a magnetic field, there is a "stiffening" of at least a part of the spin glass during the waiting time. This "stiffening" contributes to the free energy and is associated with the maximum energy barrier, height, $\Delta(t_w)$, and growth during the waiting time. This energy barrier controls the decay of the waiting time-dependent magnetization through the Arrhenius law, Equation 10, [40, 41].

$$\delta f_{SG} = \Delta(t_w) = k_B T_m (\ln t_w - \ln \tau_o). \quad (10)$$

Since the sample is held at temperature $T_m$, we take $\tau_o = \frac{h}{k_B T_m}$ as the equilibrium fluctuation time scale.

It is clear that the suppression of the waiting time effect is already in effect at a lower temperature (below the crossover line), as observed in the reduction of the peak times $t_w^{eff}$ (from $2t_w$). Figure 8B shows that the suppression begins at a temperature at least as low as $0.8T_g$. We propose that this reduction of $t_w^{eff}$ is caused by the onset of significant critical fluctuations. This region is far away from the spin glass transition temperature, and therefore, critical fluctuations will be small (size scale) and of short duration having small yet noticeable effects on $t_w^{eff}$. As the crossover line is approached, critical fluctuations become larger and are of longer duration, increasing their effect on $t_w^{eff}$. The crossover line represents a crossover from a state that can maintain rigid spin glass clusters (up to the crossover time scale) to one dominated by large long-lived critical fluctuations that suppress the ordering associated with the waiting time.

In particular, at the crossover line, the small spin glass energy associated with the waiting time effect is

$$\delta f_{SG}(t_{co}) = \Delta(t_{co}) = k_B T_{co} (\ln t_{co} - \ln \tau_o). \quad (11)$$

We would expect the free energy perturbation to be a continuous function through the crossover region.

$$\delta f_{SG}(t_{co}) = \delta f_{critical}(t_{co}), \quad (12)$$

where $\delta f_{critical}(t_{co})$ is the free energy available, within the spin glass remanence, as the crossover line is crossed, and critical fluctuations dominate the system.

We expect that the magnetization decay in the critical region would be governed by the critical fluctuation time scale $\tau_c$, which exhibits a critical decrease as the transition temperature $T_c$ is approached.

$$\tau_c = \tau_{oc} \left| \frac{T - T_c}{T_c} \right|^{-z\nu}. \quad (13)$$

We propose an effective Arrhenius law in the critical region

$$\delta f_{critical}(t_{co}) = \Delta(t_{co}) = k_B T_{co} (\ln A - \ln \tau_c). \quad (14)$$

In this equation, we assume that $\tau_{oc}$ is governed by the fixed point at $T_c$ and, therefore, $\tau_{oc} = \frac{h}{k_B T_c}$. At this stage in the analysis, $A$ is an unknown parameter with units of time and will be used as a fitting parameter.

Substituting Equation (11) and (14) in Equation 12 leads to

$$\frac{t_{co}}{\tau_o} = \frac{A}{\tau_{oc}}. \tag{15}$$

Replacing $\tau_o$ and $\tau_{oc}$ with the expressions defined above leads to

$$t_{co} = \frac{AT_c}{T_{co}\left|\frac{T-T_c}{T_c}\right|^{-z\nu}}. \tag{16}$$

Although this equation can be used to fit the data for $t_{co}$ vs. T with different $A$ values, we seek to understand the magnetic field effect on the crossover lines. It is clear from Figure 8 that the magnetic field suppresses $t_{co}$. We, therefore, write

$$t_{co} = \frac{AT_c}{g(H)T_{co}\left|\frac{T-T_c}{T_c}\right|^{-z\nu}}, \tag{17}$$

where $g(H)$ is to its first order a monotonically increasing function. The first attempt to find $g(H)$ was to bring the magnetic field in by adding a Zeeman term in Equation 10 [42]. This was first used to probe the spin glass energy barrier landscape with a magnetic field. We find that adding this field dependence into Equation 10 and deriving through Equation 18 yield an expression for g(H) that is an exponential of $H^2$. With this field dependence, we cannot fit the data. Likewise, using the scaling ansatz for the critical region also brings an exponential of $H^{\frac{2}{\delta}}$ into Equation 16, and using an accepted value of $\frac{2}{\delta} = .216$, we also cannot fit the data.

We do find, however, that the data fit to $g(H) = \alpha H^2$. With the substitution $A' = \frac{A}{\alpha}$, we fit our data to

$$t_{co} = \frac{A'T_c}{H^2 T_{co}}\left|\frac{T-T_c}{T_c}\right|^{z\nu}. \tag{18}$$

Figure 9 is a plot of Equation 18 using $\tau_{oc} = \frac{h}{k_B T_c} = 1.48 \times 10^{-12}$ s and ($z\nu$ = 7) [10]. With two fitting parameters, $T_c$ = 32.4 K (1.03$T_g$) and $A'$ = 3 $Oe^2s$ ($3 \times 10^{-8}$ $T^2 s$), we find that Equation 18 fits to the 9.6, 16, 32, 64, and 96-Oe crossover lines. We do not include the 16-Oe data in this plot, but we include the fit of Equation 18 with H = 16 Oe. We also plotted Equation 18 for H = 16 Oe in Figure 8B (the solid black line). It can be observed in Figure 8B that Equation 18 fits well above 29 K, but below 29 K, for $t_w$ = 100,000 s, the data, while suppressed ($t_w^{eff} \ll t_w$), are above the line. This suggests that critical fluctuations, while present and important, are not dominant below 29 K in 16 Oe.

The $H^2$ field dependence is strongly suggestive that the suppression of $\tau_{oc}$ by the magnetic field may be caused by the magnetic susceptibility (i.e., $g(H) \propto \chi_{SG}$). In particular, as $T_c$ is approached, it is expected that the nonlinear terms in the magnetization expansion diverge. The expansion of the susceptibility leads to [10, 43]

$$\chi_{SG} = \chi_1 + \chi_3 H^2 + \chi_5 H^4 + \chi_7 H^6 \ldots, \tag{19}$$

where the nonlinear terms $\chi_3$, $\chi_5$, $\chi_7$ etc., are expected to diverge as $T_c$ is approached. We analyzed the above data with Equation 19. However, with only five fields, polynomial fits to the $H^4$ and $H^6$ are problematic. Recently, we extended our measurements over the above temperature region with up to 11 different magnetic fields per temperature. It can also be observed in Figure 9 that as $T_c$ is approached, the higher-field data break off from Equation 18 and approach the lower-field data. This is also likely a nonlinear effect, and the analysis is forthcoming [44].

Finally, we discuss an observation made with respect to the field dependent data and some open questions. In Figure 8, it can be observed that as the magnetic field decreases, the crossover line becomes more vertical and shifts to higher temperatures. The line at 28 K is an aid to the eye. This clear difference implies that as $H \to 0$ G, the crossover line may become vertical at the transition temperature. Does this imply that the coherence length of the ZFC-TRM can grow without bounds, even close to $T_c$? This shift also explains the breakdown of superposition between the FC-TRM and the ZFC-TRM [19]. In the ZFC-TRM, the spin glass is cooled in zero field and ages during the waiting time without the field-dependent suppression observed in the crossover line. The FC-TRM ages with the field-dependent suppression. The data and analysis in this paper strongly imply a phase transition at H = 0 Oe. In a magnetic field, the situation is not quite so clear. Does a limit on the coherence length, as observed in a magnetic field, imply a crossover transition in a magnetic field? Further understanding of the role of the coherence length and its intrinsic role in the spin glass state is warranted.

In summary, we measured the TRM decays for a range of magnetic fields and temperatures below the transition temperature. We find that on the approach to $T_c$, the waiting time effect is suppressed and reaches a cut-off time scale $t_{co}$. We mapped this suppression over five magnetic fields, from 9.6 to 96 Oe, and find that $t_{co}$ is a function of the magnetic field. We postulate that the free energy of this effect is a continuous function through $t_{co}$. Using the Arrhenius law as the governing mechanism for the decay of the TRM, we understand the cut-off in terms of critical dynamics. In a previous paper [33], lead by the apparent upward curvature of the crossover line, we fit the crossover line at 16 G to both a glass transition and a low temperature phase transition. In the Supplementary Material, we extend this analysis to the data for the magnetic fields reported in this paper.

## Data availability statement

The raw data supporting the conclusion of this article will be made available by the authors, without undue reservation.

## Author contributions

GK: conceptualization, data curation, formal analysis, funding acquisition, investigation, methodology, project administration, resources, software, supervision, validation, visualization, writing–original draft, and writing–review and editing. MB: data curation and writing–review and editing. RB: data curation and writing–review and editing. MH: data curation and writing–review and editing. DT: data curation, investigation, and writing–review and editing.

## Funding

The author(s) declare that financial support was received for the research, authorship, and/or publication of this article. This work was supported by the US Department of Energy (USDOE), Office of Science, Basic Energy Sciences, under Award DE-SC0013599. The IUP Dual DC SQUID magnetometer was built under an NSF MRI, Award No. 0852643. Single-crystal growth was performed by Deborah Schlagel at the Ames Laboratory, which is supported by the Office of Science, Basic Energy Sciences, Materials Sciences and Engineering Division of the USDOE, under Contract No. DE-AC02-07CH11358.

## Acknowledgments

The authors thank R.L. Orbach, J. Meese, P. Young, E.D. Dahlberg, and J. Friedberg for useful discussions. The authors also thank GK for his help with the manuscript.

## Supplementary material

The Supplementary Material for this article can be found online at: https://www.frontiersin.org/articles/10.3389/fphy.2024.1443298/full#supplementary-material